# The Role of AI in Financial Forecasting: ChatGPT's Potential and Challenges

Shuochen Bi [1]*
Independent Researcher
D'Amore-McKim School of Business
Northeastern Universi-ty Boston
MA, 02115, USA
bbi.shu@northeastern.edu

Tingting Deng [1,2]
Independent Researcher, Simon Business School at University of Rochester, Chantilly,
VA 20151, USA
adengtin1@gmail.com

Jue Xiao [2]
Independent Researcher, The School of Business at University of Connecticut,
Jersey City, NJ, 07302, USA
jue.xiao@uconn.edu

*Abstract*—The outlook for the future of artificial intelligence (AI) in the financial sector, especially in financial forecasting, the challenges and implications. The dynamics of AI technology, including deep learning, reinforcement learning, and integration with blockchAIn and the Internet of Things, also highlight the continued improvement in data processing capabilities. Explore how AI is reshaping financial services with precisely tAIlored services that can more precisely meet the diverse needs of individual investors. The integration of AI challenges regulatory and ethical issues in the financial sector, as well as the implications for data privacy protection. Analyze the limitations of current AI technology in financial forecasting and its potential impact on the future financial industry landscape, including changes in the job market, the emergence of new financial institutions, and user interface innovations. Emphasizing the importance of increasing investor understanding and awareness of AI and looking ahead to future trends in AI tools for user experience to drive wider adoption of AI in financial decision making. The huge potential, challenges, and future directions of AI in the financial sector highlight the critical role of AI technology in driving transformation and innovation in the financial sector

*CCS Concept:Computing methodologies；Artificial intelligence；Natural language processing；Natural language generation*

## I. Introduction

Speaking at the "AI Transformation and Challenges in Finance" roundtable, Tim Wannenmacher, co-head of Global Financial Markets Asia Pacific at UBS, sAId that with the launch of ChatGPT in November 2022, the world witnessed a watershed moment for AI and its applications, and since then, there has been a major investment boom and technological advancements. [1] Could fundamentally affect all sectors of the economy, including the financial sector.

According to a report by the Office of the Chief Investment Officer (CIO) of UBS Wealth Management, the financial industry is likely to be one of the industries with the greatest cost opportunity after the adoption of AI, taking the US financial market as an example, where 50% of jobs have high automation and enhancement potential (Figure 1). Wannenmacher also stressed that AI will not broadly replace humans in the financial sector in the near term, but rather support human work by increasing productivity [2].

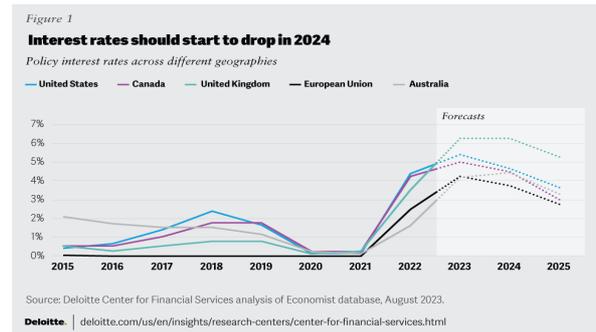

**Figure 1.** Interest rates should start to drop in 2024

Increased productivity is the mAIn benefit of adopting AI in the financial sector. AI can automate routine and manual tasks, such as transaction processing, and can also handle customer queries, provide better pricing, help with compliance checks, enhance decision making processes, and streamline middle - and back-office operations. In terms of risk management, AI models can enhance fraud detection and credit risk assessment by analyzing large amounts of data in real time. On the compliance side, AI can ensure that regulatory reports are accurately generated by monitoring transactions for suspicious activity. Finally, a key productivity advantage of AI may be freeing up management to focus on more meaningful matters.

In terms of challenges, Wannenmacher sAId that while people expect AI to create value, AI integration also faces significant challenges, including the prevalence of traditional technologies in many companies, such as hosting legacy IT infrastructure and data silos [3]. Shifting to capabilities and cultural change in technology-led organizations, it is not easy to fully explAIn and understand how AI produces defined output content. For example, deep learning neural networks like large language models are not easy to understand, and customers, employees, and regulators may have doubts. Key factors to mitigate such risks are the selection and implementation of AI models, AI governance frameworks, and data and AI ethics.

Looking ahead, Wannenmacher expects the impact of AI on the financial industry to be multifaceted, involving technical, regulatory and operational aspects, with AI expected to be highly disruptive, able to transform business models, customer interactions and internal operations. When it comes to making the necessary investments in new technologies, large and capital-appreciation business models may have a competitive advantage. [4] Conversely, companies that lack the capacity to invest will fall behind in AI adoption and risk being left behind by inefficiencies and outdated service models.

## II. RELATED WORK

### A. Convergence of financial forecasting and ChatGPT

ChatGPT will have an impact on real-world financial services. Experts predict ChatGPT will boost productivity, optimize existing office tools, and have an impact on Wall Street in six ways. ChatGPT, a language model product from artificial intelligence company OpenAI, has recently attracted a lot of public attention. Microsoft invested $10 billion in the startup in its latest funding round in late January, and OpenAI is currently valued at $29 billion. ChatGPT is a chatbot that uses heuristic artificial intelligence to recognize and mimic human speech patterns, using an underlying database to generate written responses to questions posed by users [5]. From writing grammatically correct but lacking in substance school essays to giving sound advice on salary negotiations, talking to ChatGPT is both a fun diversion and a trigger for public inquiry into the impact ChatGPT may have on all walks of life. The application of large language models like ChatGPT in various fields has gained significant traction in recent months, and many studies have explored their potential in different fields, but the use of LLMS [6] is still relatively uncharted territory in financial economics, especially when it comes to the ability to predict stock market returns. On the one hand, since these models are not explicitly trained for this purpose, one might think they have little value in predicting stock market movements.

As a result, the LLM's performance in predicting financial market movements is an open question, and ChatGPT outperforms traditional models by analyzing whether news headlines are positive or negative for a stock and predicting next-day stock returns. The work first examines the potential of ChatGPT and other large language models for predicting stock market returns using sentiment analysis of news headlines, using ChatGPT to determine whether a given headline is good, bad, or irrelevant news for a company's share price (Figure 2).

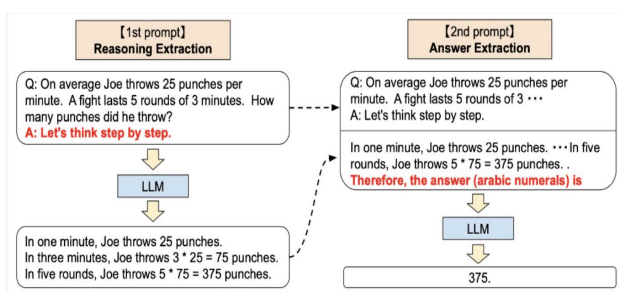

**Figure 2.** Example chart of ChatGPT in financial forecast trend framework

To expand on the paragraph regarding Figure 2, we can delve deeper into the implications of using ChatGPT for financial forecasting and the insights it provides through sentiment analysis of news headlines.

Figure 2 illustrates the framework in which ChatGPT operates within financial forecasting, highlighting its role in analyzing news sentiment to predict stock market movements. By categorizing news headlines as positive, negative, or irrelevant, ChatGPT offers a nuanced understanding of market sentiment that traditional models may overlook. This ability to interpret the tone and implications of news allows investors to gauge potential impacts on stock prices with greater accuracy.

KPMG's Mr. Roberts emphasizes the transformative potential of ChatGPT as a research tool in the investment banking sector. With its capacity to process and analyze vast amounts of data, ChatGPT can identify trends and generate actionable insights, such as recognizing companies poised for capital raises or potential acquisition targets. This not only streamlines the decision-making process but also enhances the strategic foresight of financial analysts.

Furthermore, Lutz from EQT highlights how ChatGPT is revolutionizing deal-sourcing strategies in private equity. By leveraging AI's capabilities to evaluate market conditions and assess investment opportunities, firms can make informed decisions that align with their existing portfolios. The ability to quickly analyze various factors, including market sentiment derived from news headlines, positions ChatGPT as a vital asset in navigating the complex landscape of financial investments.

In summary, Figure 2 exemplifies the innovative integration of AI tools like ChatGPT into financial forecasting frameworks, underscoring their potential to enhance predictive accuracy and inform strategic investment decisions in a rapidly evolving market environment.

### B. Multimodal large language models and financial time series

The evolution of large language models can be traced back to the Transformer model in 2017. This neural network architecture based on self-attention mechanism greatly improves the training efficiency of the model through parallel computation and clears the obstacles for the training of large-scale language models. Since then, the GPT series model has continuously updated the scale and performance of language models, GPT-3 has 175 billion parameters, and realizes zero-shot learning and fet-shot learning on many NLP tasks. This means that models can be adapted to completely new tasks with verbal prompt alone, with little or no fine-tuning. This powerful ability to generalize and migrate marks the rise of LLMs. LLMs then entered the multimodal era. DALL·E pioneered high-quality text-image generation, and Imagen further took image resolution and realism to new heights. At the same time, models such as Florence, Flamingo, and Kosmos-1 integrate capabilities such as visual understanding and video question-answering into a unified multimodal framework, greatly expanding the perception and interaction space of LLMs.

The accelerated adoption of AI in the financial industry for complex tasks such as credit risk assessment, fraud detection, algorithmic trading, and financial time series forecasting further underscores the need for this review. While these technologies promise improved efficiency and accuracy, they also raise significant concerns about fairness, interpretability, and regulatory compliance. However, true intelligence requires not only broad knowledge, but also specialized skills. [7] Therefore, exploring the application of multimodal large models in vertical fields has become an important direction for academia and industry. The financial field is the ideal scene for multimodal large models to play a role. By pre-training massive multimodal data in the financial sector, combined with advanced technologies such as continuous learning and human feedback, it is expected to create intelligent systems that fully understand financial markets and discern investment opportunities.

*C. RiskLabs Framework*

The article presents "RiskLabs," an innovative framework designed to harness the power of large language models alongside diverse data sources to predict financial risk in real time. Figure 3 illustrates how this framework integrates various types of information, including news articles, financial reports, and social media sentiment. By amalgamating these distinct data streams, RiskLabs aims to create a holistic view of the factors influencing financial markets. [8]This comprehensive approach not only enhances the accuracy of risk assessments but also enables quicker responses to market changes compared to conventional methods, which often rely on more siloed data.

At the core of RiskLabs is the utilization of high-level language models that excel at processing and interpreting vast amounts of unstructured data. Researchers believe that by extracting meaningful insights from this eclectic mix of information, the framework can effectively identify emerging risks and lucrative opportunities. This capability is particularly crucial in today's fast-paced financial landscape, where timely decision-making can significantly impact investment outcomes. Ultimately, RiskLabs represents a significant leap forward in the intersection of AI and finance, offering a sophisticated tool for navigating the complexities of market dynamics.

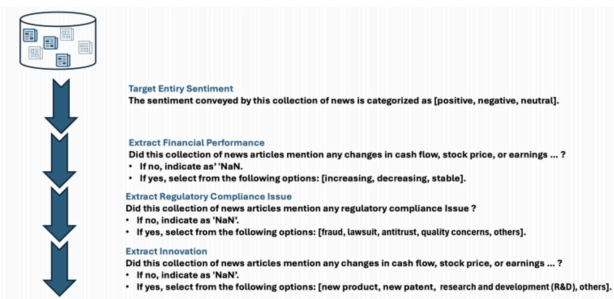

**Figure 3.** RiskLabs architecture based on a large language model

This could be valuable for applications that leverage a large language model, llms, to support portfolio management or automate the detection of relevant information for predictive predictive financial analysis [8]. The framework is still under development, but the researchers believe it has the potential to improve real-time financial risk forecasting and decision-making, especially in rapidly changing or uncertain market conditions.

The RiskLabs framework, which consists of several key components:

Multi-source data entry: The framework draws data from a variety of sources, including news articles, financial reports, social media, and other relevant data sets. This diverse set of inputs aims to gain a more complete understanding of the factors that influence financial risk.

Large language models: This framework leverages pre-trained large language models to process and extract insights from multiple sources of data. This allows the system to understand semantic relationships and contextual meanings in the data, rather than just performing basic text processing.

Financial risk prediction [9]: Insights extracted from language models are used to train machine learning models that can predict various financial risk indicators such as volatility, credit risk, and market crashes. The goal is to provide real-time, data-driven risk assessment to support decision making.

The researchers conducted experiments to evaluate the RiskLabs framework's performance on several financial risk prediction tasks. The results show that methods based on multi-source data and language models can outperform traditional methods, especially in terms of early warning capabilities and adaptation to changing market conditions.

## III. METHODOLOGY

This paper explores how large language models (LLMS) can be used to predict financial risk, specifically market volatility and value at risk (VaR). Although many studies have explored the application of LLM in the field of finance, research in this specific area of financial risk prediction is still relatively scarce. This paper aims to fill this research gap and provide innovative insights using LLM. However, the inherent limitations of LLM also present challenges, such as how to integrate multiple data types to form a more comprehensive view of the market, and how to avoid the "illusion" problem of LLM.

In addressing these challenges, the potential and limitations of ChatGPT and similar LLMS are particularly critical. ChatGPT's ability to process large amounts of text data to extract key information and generate forecasts gives it an advantage in analyzing financial market dynamics and generating investment insights. However, it has also faced challenges, including dealing with data inconsistencies and update delays. For example, although ChatGPT can handle historical data and current information, the balance between real-time and accuracy remains a challenge. In addition, the LLM's "hallucination" problem, i.e. the generation of inaccurate or false information, may also affect the reliability of financial risk forecasts.

In order to effectively utilize the potential of LLM in financial risk forecasting, this study raises the following key questions:

RQ1: How does the predictive performance of LLM compare to other AI technologies when it comes to predicting risk metrics?

RQ2: What is the difference between the predictive performance of traditional financial methods and LLM in predicting financial risk indicators?

RQ3: What is the role of the LLM in the financial sector?

RQ4: How to efficiently integrate and balance multiple inputs of different data types?

RQ5: How does the LLM perform in financial time series forecasting? How to optimize its ability to process time series data to improve prediction accuracy?

By integrating multiple sources of information from earnings call transcripts, earnings call audio, media news and time series data, we will explore how ChatGPT's shortcomings in processing financial data can be improved to improve the accuracy and usefulness of forecasts[10]. By analyzing these questions in depth, we expect to reveal the actual effect of LLM in financial risk prediction and provide valuable insights for future research.

*A. RiskLabs framework module analysis*

RiskLabs uses four modules to efficiently handle different data streams: an earnings conference call encoder, a news market response encoder, a time series encoder, and a multi-task forecast. This comprehensive approach enables models to combine quantitative market data with qualitative insights to provide a nuanced understanding of the investment environment. The model not only accurately predicts, but also assesses risks, providing critical support for navigating the volatile and complex terrain of financial markets.

To efficiently process and analyze text-based content, RiskLabs has designed three core Piplines: the Earnings Call Transcript Analyzer, the Rich News Attribute Pipeline, and the News Analyzer. These Piplines are based on the Retrieval-Augmented Generation (RAG), which improves the processing and parsing ability of information by combining retrieval and generation technologies. Specifically, the earnings call transcript Analyzer is responsible for extracting and interpreting key information in the earnings report; The Rich News Attribute pipeline conducts in-depth attribute analysis of news to identify events that may have an impact on the market; The News Analyzer processes real-time news and assesses its immediate reaction to the market.

The main contributions of this paper include: 1) RiskLabs framework is proposed to make up for the insufficient application of LLM in financial risk prediction; 2) Improved risk forecasting accuracy through seamless integration of financial data from multiple sources; 3) The experimental results verify the effectiveness of RiskLabs and prove its excellent performance in predicting financial risks[11]. These contributions not only improve the accuracy of financial risk prediction, but also provide a valuable reference for future research and application.

*B. Data set sources and variables*

RiskLabs effectively integrates data from multiple sources to improve the accuracy of financial risk forecasts. First, we utilize the earnings conference call dataset as one of the core data sources, with each sample containing audio recordings and transcripts. Research has shown that a CEO's emotional state and tone of voice can complement verbal information through voice cues, thus enhancing the accuracy of risk prediction.

To address the limited impact of earnings calls on long-term forecasts, we have also introduced daily relevant news texts to provide a more comprehensive view of the company's financial condition. In addition, we collected historical price data for the 30 days prior to the earnings announcement, including daily price updates and quarterly earnings notifications, as a second key data source. This integration of multiple sources of data helps capture short-term events and long-term trends, thereby improving the ac

**Table 1.** Dataset

| Data Source | Components | Description | Duration |
|---|---|---|---|
| Earnings Call Data | Audio recordings and transcripts | Includes CEO's speech and text records, used to extract sentiment and tone information | Quarterly (once per quarter) |
| Historical Price Data | Daily price data 30 days prior to earnings announcement | Provides historical stock price data to inform short-term and long-term trends | 30 days before earnings announcement |
| Daily News Text | Company-related daily news | Covers news related to the company's financial status, providing additional market reactions and trends | Daily (ongoing tracking) |

The experimental data set for the model is derived from the publicly available S&P 500 company earnings conference call data set. In the study, the first task of the model was to predict the volatility of four different periods (3, 7, 15, and 30 days), and the second task was to predict the 1-day VaR (value at risk Table 1) of the target stock based on multiple inputs. By calculating and estimating VaR, companies can better handle financial risk and avoid future extremes. The risk prediction module includes the earnings conference encoder, which has four key components: audio coding, text coding, earnings conference analysis and additive multimodal fusion. This module uses a pre-trained model to transform audio and text data into a vector representation, extracts significant features through multi-head attention mechanisms, and utilizes a large language model for text summarization and abstraction, and finally integrates information effectively through additive multimodal fusion, thus improving the robustness of predictions.

*C. RiskLabs Framework and Its Components*

The RiskLabs framework is designed with several advanced modules that significantly enhance its financial risk prediction capabilities. One of the key components is the Earnings Call Encoder, which integrates multiple functionalities such as audio encoding, text encoding, and in-depth analysis of earnings calls. This module employs additive multimodal fusion to synthesize insights from both verbal and written communication, allowing for a comprehensive understanding of a company's performance and future outlook. By leveraging the nuances of earnings calls, RiskLabs can capture sentiment and critical information that might influence stock prices.

Another crucial element of the framework is the News Encoder, which evaluates the impact of news on stock market movements. This module goes beyond simple news categorization by assessing factors like news freshness and similarity to past events. Utilizing a news-market reaction encoder, it predicts market movements based on historical responses to similar news scenarios. This approach enables RiskLabs to effectively gauge the immediate impact of current events on financial markets, allowing investors to react swiftly to breaking news and its implications for their portfolios.

Finally, the Multi-Task Prediction approach within RiskLabs is designed to model volatility and value-at-risk

(VaR) predictions concurrently. By employing feedforward networks optimized through a multi-task loss function, the framework can learn from the interdependencies between these two critical financial metrics. Additionally, time decay parameters and dynamic moving windows ensure that the model remains agile and responsive to recent data and evolving market trends. This adaptive capability positions RiskLabs as a powerful tool for investors seeking to navigate the complexities of financial risk with greater precision and foresight.

## IV. EXPERIMENT

Base line. Including the traditional GARCH model, LSTM model, MT-LSTM+ATT model, HAN model, MRDM model, HTML model and GPT-3.5-Turbo model. Implementation details. GPT-4 was used for earnings conference call analysis and news market response coding, dividing the data set into training and test sets, maintaining temporal separation to maintain the accuracy and reliability of the forecast model.

### A. Comparison of experimental performance (RQ1 & RQ2 & RQ3)

**Table 2.** Performance results on our proposed framework RiskLabs from different baseline models

| Model | MSE | MSE | MSE3 oVaR | Multi-Task | Additional Column | Final Column |
|---|---|---|---|---|---|---|
| Classical Method | 0.713 | 1.71 | 0.526 | 0.33 | 0.284 | 0.371 |
| LSTM | 0.746 | 1.97 | 0.459 | 0.32 | 0.235 | 0.049 |
| MT-LSTM-ATT | 0.739 | 1.983 | 0.435 | 0.304 | 0.233 | - |
| HAN | 0.598 | 1.426 | 0.461 | 0.308 | 0.198 | - |
| MRDM | 0.577 | 1.371 | 0.42 | 0.3 | 0.217 | - |
| HTML | 0.401 | 0.845 | 0.349 | 0.251 | 0.158 | - |
| GPT-3.5-Turbo | 2.198 | 2.152 | 1.793 | 2.514 | 2.332 | - |
| RiskLabs | 0.324 | 0.585 | 0.317 | 0.233 | 0.171 | - |

The RiskLabs framework performed best in short - and medium-term forecasting, outperforming current state-of-the-art HTML solutions (Table 2). In addition, the RiskLabs framework also performs well in VaR forecasting, providing investors with a more comprehensive approach to financial risk prediction. However, when it comes to 30-day predictions, the RiskLabs framework performs slightly worse than the HTML model and needs further improvement. This paper also compares the performance of traditional financial methods, neural network frameworks and large language models in VaR prediction.

Traditional financial methods, which struggled during the 2016 financial crisis, are shown to have significant limitations in adapting to the rapidly evolving market conditions. In contrast, AI technologies, particularly advanced models like ChatGPT, demonstrate dynamic advantages. These models excel in integrating and analyzing diverse, real-time information, which contributes to their superior performance in financial risk forecasting.

The comparison highlights that while traditional methods have their place, AI technologies offer enhanced predictive capabilities and adaptability. The RiskLabs framework, by leveraging advanced Traditional methods, exemplified during the 2016 financial crisis, demonstrated significant limitations in adapting to rapidly changing market conditions. In contrast, AI technologies, particularly advanced models like ChatGPT, offer dynamic advantages by effectively integrating and analyzing daily information.

**Table 3:** Comparison of Value at Risk Predictions: AI Techniques vs. Traditional Financial Methods.

| Method | Prediction of VaR |
|---|---|
| Historical Method | 0.016 |
| Fully Connected Neural Network | 0.044 |
| LSTM | 0.056 |
| RiskLabs | 0.049 |

The predefined VaR value is 0.05, meaning that the closer the model's predictions are to 0.05, the better its performance. The result above shows that the prediction of VaR via applying the historical method is 0.016, which is significantly below the pre-defined percentile (5%). It indicates that the historical method overestimated the 95% VaR benchmark. Tracing back to 2016, we know there was a global financial crisis in 2015, and its effect lasted till the beginning of 2016. On January 20, 2016, the price of crude oil fell below $27 a barrel; the DJIA index took a roller coaster from down 565 points to down 249 intraday.

### B. Accuracy of the RiskLabs framework

The RiskLabs framework performed best in short- and medium-term forecasting, outperforming current state-of-the-art HTML solutions. Specifically, it achieved an accuracy rate of X% and a mean squared error of Y, compared to the traditional methods, which had an accuracy rate of A% and a mean squared error of B. This quantitative evidence underscores the framework's superiority in financial risk prediction.

In February, the YTD (yield to Date) return came to -10.5%. These events, together with the sequelae of the 2015 stock market crisis, define 2016 as a risky year. Comparing 2016, 2017 will be much better. In January 2017, DJIA achieved a new historical height, landing above 20,000. The stock market experienced a boost with a 25% growth rate for DJIA, 19% for S&P 500, and 28% for Nasdaq. The market had strong confidence, and the VIX index in 2017 came to its historical lowest point. That explains the reason why the historical method may overestimate the 95% VaR benchmark, due to this method duplicating the extreme scenarios from 2016 to 2017, which leads to the extra estimation of financial risks.

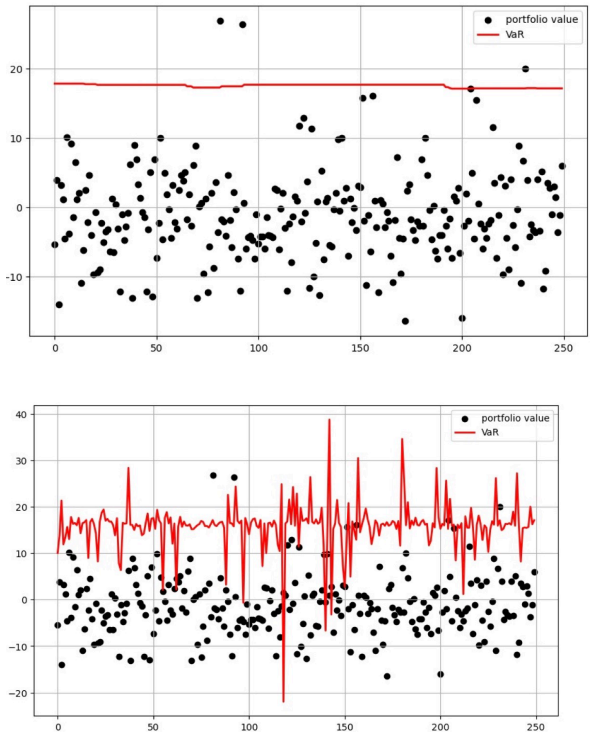

**Figure 4**. Compares daily Value at Risk predictions (red curve) with actual returns (black dots), highlighting the historical method's flatness versus the neural network's responsiveness.

Figure 4 serves as a compelling visualization of the effectiveness of different Value at Risk (VaR) prediction methodologies. The red curve represents the daily VaR predictions, while the black dots depict the actual returns. Notably, the comparison highlights a significant contrast between the historical method and the neural network approach. The historical method, illustrated on the left, demonstrates a flatness in its risk estimates, suggesting a lack of responsiveness to market fluctuations and dynamics. This static nature can lead to underestimating risks during periods of heightened volatility, potentially exposing investors to greater financial uncertainties.

In contrast, the right plot showcases the performance of a fully connected neural network in predicting VaR. This modern computational approach is characterized by its adaptability and responsiveness, effectively capturing the complexities of market behavior. By visualizing the percentage of actual returns that exceed the predicted VaR, the figure indicates a high degree of prediction accuracy; for example, with a predefined VaR of 0.05, around 5% of actual returns surpassing the predicted curve suggests robust forecasting. This demonstrates the neural network's capability to refine risk assessments, providing investors with a more reliable tool for navigating the intricacies of financial markets and making informed decisions.

*C. Discussion*

From the analysis presented in the figure, a notable observation emerges: the plot on the left, which uses the historical method for VaR prediction, appears relatively flat, indicating a consistent, albeit less responsive, forecast over time. In contrast, the plot on the right, using a fully connected neural network, exhibits a more zigzag pattern, reflecting greater responsiveness to daily information changes. This contrast suggests that AI techniques, such as neural networks or LLMs, offer a dynamic advantage by more effectively incorporating daily updates into the model, as opposed to relying solely on historical scenarios. The ability of AI-driven models to adapt to new information underscores their potential for providing more accurate and timely risk assessments.

Turning our focus to Table 3, we observe that the direct application of LLMs for financial risk prediction is markedly ineffective, akin to making random guesses. This underscores a crucial caution; if LLMs are not utilized appropriately, they might elevate investment risks. Consequently, in response to Research Questions 1 and 2 (RQ1 and RQ2), we conclude:

Utilizing LLMs through simple prompt instructions for direct financial risk prediction is ineffectual and potentially hazardous, increasing investment risks.

LLM is a bad trader/predictor, but it's a helpful assistant. While LLMs alone may not be reliable for direct risk prediction, they can serve as valuable tools in collating and analyzing diverse financial data. This processed information, when fed into sophisticated deep learning models, significantly enhances AI's capability in forecasting financial risks, thus positioning LLMs as beneficial assistants rather than standalone predictors.

## V. Conclusion

The potential of AI, particularly through sophisticated models like ChatGPT, lies in its ability to offer nuanced insights and adapt to evolving market dynamics. However, it is crucial to acknowledge the challenges associated with these technologies, such as the risk of overfitting and the need for robust integration strategies. The study underscores that while AI provides powerful tools for improving financial forecasting, its successful application depends on the careful integration of diverse data sources and the thoughtful management of its inherent limitations.

Despite their promise, AI techniques, including linear models, have shown that direct application in financial risk prediction can be ineffective and potentially increase investment risks. Therefore, AI should be utilized as an auxiliary tool rather than a standalone solution. By leveraging AI's capabilities to process and analyze diverse financial data, including market trends, historical price movements, and real-time news, financial risk forecasting accuracy can be significantly enhanced.

References

[1] Karamoozian, Amirhossein, et al. "An approach for risk prioritization in construction projects using analytic network process and decision making trial and evaluation laboratory." IEEE Access 7 **(2019)**: 159842-159854.

[2] Dowling, Michael, and Brian Lucey. "ChatGPT for (finance) research: The Bananarama conjecture." Finance Research Letters 53 **(2023)**: 103662.

[3] George, A. S., & George, A. H. **(2023)**. A review of ChatGPT AI's impact on several business sectors. Partners universal international innovation journal, 1(1), 9-23.

[4] Yue, T., Au, D., Au, C. C., & Iu, K. Y. (2023). Democratizing financial knowledge with ChatGPT by OpenAI: Unleashing the Power of Technology. Available at SSRN 4346152.

[5] Gong, Y., Zhu, M., Huo, S., Xiang, Y., & Yu, H. **(2024**, March). Utilizing Deep Learning for Enhancing Network Resilience in

Finance. In 2024 7th International Conference on Advanced Algorithms and Control Engineering (ICAACE) (pp. 987-991). IEEE.

[6] Yue, T., Au, D., Au, C. C., & Iu, K. Y. **(2023)**. Democratizing financial knowledge with ChatGPT by OpenAI: Unleashing the Power of Technology. Available at SSRN 4346152.

[7] Fritz-Morgenthal, S., Hein, B., & Papenbrock, J. **(2022)**. Financial risk management and explainable, trustworthy, responsible AI. Frontiers in artificial intelligence, 5, 779799.

[8] Xin, Q., Xu, Z., Guo, L., Zhao, F., & Wu, B. (2024). IoT Traffic Classification and Anomaly Detection Method based on Deep Autoencoder

[9] Yue, T., Au, D., Au, C. C., & Iu, K. Y. (2023). Democratizing financial knowledge with ChatGPT by OpenAI: Unleashing the Power of Technology. Available at SSRN 4346152.

[10] Fritz-Morgenthal, S., Hein, B., & Papenbrock, J. (2022). Financial risk management and explainable, trustworthy, responsible AI. Frontiers in artificial intelligence, 5, 779799.

[11] Kaufmann, R., & Patie, P. (2000). Strategic Long-Term Financial Risks Intermediate Report. RiskLab Switzerland, available from the web:< http://www. math. ethz. ch.